\documentclass[12pt,a4paper]{article}
\pdfoutput=1
\usepackage{jheppub}
\usepackage{slashed}
\usepackage{blkarray}
\usepackage{leftidx}
\usepackage[makeroom]{cancel}
\usepackage{caption}
\usepackage{subcaption}
\usepackage{comment}
\makeatletter
\def\@fpheader{\relax}
\makeatother
\usepackage{tensor}
\usepackage{graphicx}
\usepackage{amsmath,amsfonts,amssymb}
\usepackage{url}
\usepackage{mathabx}
\usepackage{tabularx}
\DeclareMathOperator{\MyProd}{\scalebox{1.4}{$\mathrm{I\kern-0.2ex I}$}}
\usepackage{mathrsfs}

\usepackage[dvipsnames]{xcolor}

\def\mO{\mathcal{O}}
\def\mD{\mathcal{D}}
\def\mF{\mathcal{F}}
\def\barz{\Bar{z}}

\def\NI{\mathscr{I}}
\def\d{\text{d}}

\title{\boldmath Subleading Asymptotic Charges in Massless Scalar QED}

\author[a]{\'{E}anna \'{E}. Flanagan}
\author[a]{ Zhihan Liu}

\affiliation[a]{Department of Physics, Cornell University, Ithaca, NY, 14853}

\abstract{We study subleading asymptotic charges of a U(1) gauge field coupled to
a charged massless scalar field in Minkowski spacetime. We consider a
phase space for which the scalar-field initial data on null infinity have compact support, while the gauge-field initial
data admit an asymptotic expansion with inverse-power tails in retarded (advanced) time, which we refer to as
initial-data tails. We identify three finite charges at the null
infinities which are conserved under scattering. One is a renormalized
version of the standard subleading soft charge. The second is a
logarithmic charge which is the difference of the coefficients of the
leading order tails at the future and past ends of null infinity. The third
is a new charge that probes the average of the tail coefficients. We
conjecture that the third charge is associated with a new class of
large gauge transformations which diverge linearly in retarded time.}


\arxivnumber{}

\begin{document}
\maketitle


\section{Introduction
\label{sec: Introduction}}

In recent years, the study of asymptotic symmetries and
charges in gauge theory has revealed a rich infrared structure even in
the simplest abelian models \cite{Strominger:2017zoo}. In
four-dimensional Maxwell theory, large gauge transformations acting
nontrivially on the celestial two-sphere 
organize an infinite set of conserved charges. Their Ward identities reproduce
soft photon theorems, and their fluxes encode electromagnetic
memory. The relations between soft theorems, asymptotic symmetries,
and memory were first uncovered in the context of gravity
\cite{Bondi:1962px,Sachs:1962wk,Strominger:2014pwa,He:2014laa}, but
have similar nontrivial realizations in QED and related abelian
theories \cite{He:2014cra,Kapec:2015ena,
  Campiglia:2015qka,He:2019pll}. 

The general form of conservation laws associated with large gauge transformations in QED is as follows. Consider a large gauge parameter $\epsilon^+(\hat{x})$ $(\epsilon^-(\hat{x}))$ on null infinity $\NI^+$ ($\NI^-)$, as a function on the unit sphere parametrized by a unit spatial vector $\hat x$ in a specific Lorentz frame. The corresponding conservation law takes the form
\begin{equation}
    Q^{+}[\epsilon^{+}]=Q^{-}[\epsilon^{-}]~,
    \label{eqn: asym-charge}
\end{equation}
provided that the large gauge parameters $\epsilon^\pm$ obey the antipodal identification 
\begin{equation}
  \label{eqn: antipodal}
    \epsilon^+(\hat{x})= \pm \epsilon^-(-\hat{x})~.
\end{equation}
The charges $Q^{\pm}[\epsilon^{\pm}]$ are evaluated at the past/future boundaries $\NI^+_-$/$\NI^-_+$ of the future/past null infinities respectively, 
and the appropriate sign in \eqref{eqn: antipodal} depends on which charge is being considered. More generally, there is an infinite tower of such conservation laws, each associated with a nontrivial conserved charge \cite{Strominger:2021mtt, Campiglia:2018dyi,Nagy:2024jua}.

The leading member of this tower is associated with Weinberg’s soft photon theorem \cite{Weinberg:1965nx}, whose form is universal and exact to all orders in perturbation theory. The subleading soft photon (and soft graviton) theorems \cite{Low:1958sn,Lysov:2014csa,Cachazo:2014fwa} are more delicate. In the quantum theory, they receive infrared-sensitive corrections, which appear as a logarithmic dependence on the soft energy in the subleading soft factor \cite{Laddha:2018myi,Sahoo:2018lxl}.  
Classically, the same logarithmic behavior is tied to the 
tail part of the gauge field, whose restriction to the generators of
$\NI^+$ contains polynomial $1/u^n$ tails
\cite{Campiglia:2019wxe,Choi:2024ygx} for all $n\geq1$, where $u=t-r$
is the retarded time, even if no such tails are present in the initial
data on $\NI^-$. The infrared limit encodes more than just the leading
soft theorem, and this suggests that, beyond the familiar leading and
subleading charges, there should be a corresponding family of
logarithmic charges sensitive to the early- and late-time tails of the
radiative data. Both leading and subleading charges have been
previously studied in Refs.\ 
\cite{Strominger:2013jfa,Kapec:2014opa,Lysov:2014csa,
  Campiglia:2014yka, He:2014cra,Campiglia:2015qka,Kapec:2015ena,
  Campiglia:2016hvg, Campiglia:2017mua,Henneaux:2018gfi}.

In this paper, we give a purely classical account of these structures
in Maxwell theory coupled to a massless charged scalar field. We
define a radiative phase space at null infinity in which the free data
for the gauge field admit early/late-time power-law
tails (\(1/u^n\)). Within this phase space, we derive explicit
expressions for the leading \cite{He:2014cra,
  Campiglia:2015qka,Kapec:2015ena} and subleading
\cite{Lysov:2014csa,Kapec:2015ena,Campiglia:2016hvg} large-gauge
charges associated with arbitrary angle-dependent parameters
\(\epsilon(\hat{x})\), and we make explicit how the \(1/u\) tails
enter these charges. We then construct a logarithmic charge whose soft
part reproduces the logarithmic contribution to the subleading soft
factor in the corresponding soft theorem, and is well defined on our
phase space. This charge coincides with the logarithmic charge defined
in the context of massive sources in Refs. \cite{Campiglia:2019wxe,
  Compere:2025tzr, Choi:2024mac, Choi:2024ygx, Hirai:2018ijc}. In addition, we
identify a new infinite family of conserved charges associated with
the \(1/u\) tails. To our knowledge, this class of charges has not
been isolated before. 

Our method of analysis is based on that of Ref.\,\cite{Campiglia:2018dyi} and is as follows. First, we analyze the asymptotic Maxwell-scalar dynamics in a neighborhood of null infinity using a systematic large-$r$ expansion ansatz. Substituting this ansatz into the field equations yields a hierarchy of constraint equations, which we solve recursively order by order in $1/r$ and $\log r/r$. This determines the asymptotic fields and allows the leading, subleading, and logarithmic large-gauge charges to be written as surface integrals over $\NI^+_-$ and $\NI^-_+$, with the contribution of the $1/u$ tails made explicit. Spatial infinity then serves as the narrow bridge across which these candidate charges are tested for genuine conservation. We therefore pass to hyperbolic coordinates near $i^0$, relate the large-$r$ expansions on $\NI^-$ and $\NI^+$ and derive the matching conditions that imply the corresponding conservation laws on the phase space. Throughout, the analysis is purely classical and complements approaches based on the covariant phase space formalism, in which divergent $\mO(r^n)$ large gauge transformations are realized canonically by constructing a renormalized symplectic form on an extended radiative phase space \cite{Campiglia:2016hvg,Choi:2024mac,Peraza:2023}.

This paper is organized as follows. In section \ref{sec: MaxwellEqn}
we review the asymptotic analysis of Maxwell's equations at null
infinity with a massless charged scalar field source, set up
the large-$r$ expansion ansatz, and define the phase space with
initial-data power-law tails. In section \ref{sec:
  Conserved-charge} we solve the resulting hierarchy of equations near
$\NI^\pm$ and derive explicit expressions for the leading and
subleading large-gauge charges, manifesting how the initial-data tails
contribute. In the same section we construct a logarithmic charge
whose soft part reproduces the logarithmic contribution to the
universal factor in the subleading soft photon theorem, and we
identify a conserved charge associated with the $1/u$ tails. In
section~\ref{sec: Matching-Condition} we study the behavior of the
fields near spatial infinity in hyperbolic coordinates, establish the
matching conditions relating the data on $\NI^-$ and $\NI^+$, and show
that the leading, subleading and logarithmic charges, together with
the new tower, satisfy the corresponding conservation laws.

\section{Asymptotic phase space for massless scalar QED}
\label{sec: MaxwellEqn}
In this section, we review and generalize the standard asymptotic expansion of Maxwell fields at null infinity \cite{Campiglia:2016hvg,Campiglia:2018dyi}. 

\subsection{Action principle and equations of motion}
\label{subsec: Action}
\indent \indent We consider a $U(1)$ Maxwell field $A_\mu$ coupled to a
massless complex scalar field $\Phi$ on four-dimensional Minkowski spacetime $M$. The action is 
\begin{equation}
    \begin{split}
        S= & -\int_{M} \d^4x\sqrt{-g}\left[\frac{1}{4e^2}F_{\mu\nu}F^{\mu\nu}+\left(D_\mu\Phi \right)^\ast\left(D^\mu\Phi\right)\right]~,
    \end{split}
    \label{eqn: Maxwellaction}
\end{equation}
where $F_{\mu\nu}=\nabla_\mu A_\nu-\nabla_\nu A_\mu$ is the field strength, and $D_\mu=\nabla_\mu+iA_\mu$ is the gauge-covariant derivative. The equations of motion are given by 
\begin{subequations}
    \label{eqn: EoM}
    \begin{align}
    \nabla_\nu F^{\nu\mu} =& -j^\mu\equiv -ie^2\left[\Phi^\ast \left(D^\mu\Phi\right)- \Phi \left(D^\mu\Phi\right)^\ast\right]~, 
    \label{eqn: EMEoM}\\
     D_\mu D^\mu \Phi=& 0~,
    \label{eqn: scalarEoM}
    \end{align}
\end{subequations}
where we follow the convention of Ref.\, \cite{Campiglia:2018dyi}. 


\subsubsection{Maxwell's equations near null infinity}
\indent \indent To understand the asymptotic symmetries, we are
interested in solutions of the field equations
(\ref{eqn: EoM})
near the null infinities $\NI^\pm$. For future null infinity $\NI^+$, the metric of Minkowski spacetime in retarded coordinates $(u,r,\theta^A)$ reads
\begin{equation}
    d s^2=-du^2-2dudr+r^2q_{AB}d\theta^Ad\theta^B~,
    \label{eqn: retard}
\end{equation}
where $q_{AB}$ is the metric on a unit two-sphere $S^2$ and we use
abstract capital Roman indices $A, B,\dots$, to refer tensors on
$S^2$.  It will also sometimes be convenient to specialize to 
holomorphic coordinates $(z,\barz)$ for which
\begin{equation}
    q_{AB}\d \theta^A\d\theta^B=2\gamma_{z\barz}\d z\d \barz=\frac{4\d z\d \barz}{(1+z\barz)^2}~.
    \label{eqn: gzz}
\end{equation}
Below we will use $\hat{x}$, $(z,\barz)$ and $\theta^A$
interchangeably to parameterize points on the two-sphere.

The equations of motion \eqref{eqn: EoM} in the coordinates \eqref{eqn: retard} become \cite{Campiglia:2018dyi}:
\begin{subequations}
    \label{eqn: Maxwell-eqn}
    \begin{align}
    \label{eqn: Maxwell-r}
    r^2j_r &=-\partial_r(r^2F_{ru}) + D^AF_{rA}~,\\
    \label{eqn: Maxwell-u}
    r^2j_u &= -\partial_r(r^2F_{ru}) + r^2\partial_uF_{ru} + D^AF_{uA}~,\\
    \label{eqn: Maxwell-A}
    j_A & = \partial_r(F_{uA}-F_{rA}) + \partial_uF_{rA} + r^{-2}D^BF_{AB}~, 
    \end{align}
\end{subequations}
where $D_A$ is the covariant derivative on $S^2$ and
$D^A=q^{AB}D_B$. The field equations near past null infinity in
advanced coordinates $(v,r,\theta^A)$ with $v=2r+u$ are similar.

As proposed by Campiglia and Laddha \cite{Campiglia:2018dyi}, it is more convenient to work with the complex self-dual and anti-self-dual fields of electromagnetism than the conventional real fields. Defining the dual field via the Hodge dual operation
\begin{equation}
    \Tilde{F}_{ab}= \frac{1}{2}\epsilon_{abcd}F^{cd}~,
    \label{eqn: dual-field}
\end{equation}
we define the self-dual (sd) and anti-self-dual (asd) combinations as
\begin{equation}
    F^{\rm sd}_{ab}=F_{ab}- i\Tilde{F}_{ab}~,\quad F^{\rm asd}_{ab}=F_{ab}+ i\Tilde{F}_{ab}~.
    \label{eqn: self-dual field}
\end{equation}
We will focus on the chiral sector $F^{sd}_{ru}$, which corresponds to quantized photon states with negative helicity. The corresponding charges have real electric and imaginary magnetic parts.

From Maxwell's equations \eqref{eqn: EMEoM} and the definition
\eqref{eqn: self-dual field} of the
self-dual field, we obtain
\begin{subequations}
    \begin{align}
        \label{eqn: Maxwell-self-dual}
        \nabla^aF_{ab}^{\rm sd}=-j_b~,\\
        \label{eqn: sd-def}
        \Tilde{F}_{ab}^{\rm sd}=iF_{ab}^{\rm sd}~.
    \end{align}
\end{subequations}
The explicit form of Eq.\ \eqref{eqn: Maxwell-self-dual} in the
retarded coordinates $(u,r,\theta^A)$ is given by replacing $F_{ab}$ in
Eqs.\ \eqref{eqn: Maxwell-eqn} by $F^{\rm sd}_{ab}$.  Combining the
resulting equations yields
\begin{equation}
    2D^zj_z=-\partial_r \left[r^2j_r + \partial_r \big(r^2F_{ru}^{\rm sd} \big) \right] + 2\partial_u \left[r^2j_r + \partial_r \big(r^2F_{ru}^{\rm sd} \big) \right] -\Delta  F_{ru}^{\rm sd}~,
    \label{eqn: j-Fru}
\end{equation}
where $\Delta =q^{AB}D_AD_B$ is the Laplacian on the two-sphere $S^2$.
Equation (\ref{eqn: j-Fru}) will be sufficient for most of our analyses
near $\NI^+$.

\subsection{Definition of phase space}

\indent\indent In this subsection we define the phase space in which our analysis will be carried out by specifying a space of free data on $\NI^+$ for the coupled Maxwell-scalar system, and defining a parameterization that
completely fixes the gauge freedom.  

Near future null infinity, we consider fields admitting the asymptotic expansions
\begin{subequations}
    \label{eqn: Asymp-fields}
    \begin{align}
    A_r(r,u,\hat{x}) &= \mO(r^{-2})~, \label{eqn: Ar-falloff}\\
    A_u(r,u,\hat{x}) &= \overset{0}{A}_u(u,\hat{x}) + \mO(r^{-1})~, \label{eqn: Au-falloff}\\
    A_A(r,u,\hat{x}) &= \overset{0}{A}_A(u,\hat{x}) + \mO(r^{-1})~, \label{eqn: AA-falloff}\\
    \Phi(r,u,\hat{x}) &= \sum_{n=1}\frac{\overset{n}{\phi}(u,\hat{x})}{r^{n}}~.
    \label{eqn: Phi-falloff}
    \end{align}
\end{subequations}
Following Refs.\ \cite{Strominger:2017zoo,Satishchandran:2019pyc}, we use the gauge freedom to impose
\begin{subequations}
    \label{eqn: gauge-choice-Iplus}
    \begin{align}
        A_r(r,u,\hat{x})=&0~,\\
        \overset{0}{A}_u(u,\hat{x})=&0~.
        \label{eqn: gauge-fix-Au}
    \end{align}
\end{subequations}
Note that Eq.\ \eqref{eqn: gauge-fix-Au} is an asymptotic gauge
condition that constrains the fields only near $\NI^+$, not everywhere
in spacetime. In this gauge, the outgoing radiative data on $\NI^+$
are given by the pair 
\begin{equation}
    \Big\{\overset{0}{A}_A(u,\hat{x}),\,\overset{1}{\phi}(u,\hat{x})\Big\}~.
    \label{eqn: free-data}
\end{equation}
The residual gauge freedom acts as
\begin{equation}
\overset{0}{A}_A \mapsto \overset{0}{A}_A + D_A\epsilon~, \qquad \overset{1}{\phi} \mapsto e^{-i\epsilon}\overset{1}{\phi},
\end{equation}
where the gauge parameter $\epsilon=\epsilon(\hat{x})$ is a function on $S^2$ only. Once these data are specified, the coupled equations of motion \eqref{eqn: EoM} determine $A_\mu$ and $\Phi$ everywhere in spacetime up to gauge. Hence, the phase space is parametrized by the data (\ref{eqn: free-data}) on $\NI^+$.

We further restrict the space of free data to configurations with sufficiently regular behavior at early and late retarded times. Specifically, we impose the following large-$|u|$ behavior:
\begin{subequations}
\label{eqn: regular}
  \begin{align}
        \label{eqn: regularity-condition-A}
        \overset{0}{A}_A(u,\hat{x})=
        &\,\sum_{n=0}\frac{\overset{0,n}{A^\pm_A}(\hat{x})}{u^n} =
        \overset{0,0}{A^\pm_A}(\hat{x})+\frac{\overset{0,1}{A^\pm_A}(\hat{x})}{u}+\dots~,\quad
        u\to \pm\infty\\
                \label{eqn: regularity-condition-B}
    \overset{1}{\phi}(u,\hat{x})= &{\rm \ compact\ support\ in\ } u.
    \end{align}
    \label{eqn: regularity-condition}
\end{subequations}
The expansion (\ref{eqn: regularity-condition-A})
is to be understood in the asymptotic sense, in particular no
assumption of convergence is being made.   
Note that these fall-off conditions (\ref{eqn: regular}) are assumptions that define the
phase space, not consequences of the field equations.  When the
conditions are satisfied, the field configurations admit finite
Poincar\'e charges and a finite symplectic form\footnote{It is possible to
specify more general regularity conditions that achieve these goals,
but the subsequent analysis would then be more involved.}. 

We similarly define the space of incoming free data
$\Big\{\overset{0}{A}_A(v,\hat{x}),
\overset{1}{\phi}(v,\hat{x})\Big\}$ on $\NI^-$ with regularity
conditions similar to the conditions \eqref{eqn: regularity-condition}. Assuming that
these two function spaces are compatible under scattering, we define
the phase space $\mathscr{F}$ to be the space of free data \eqref{eqn:
  free-data} satisfying the fall-off conditions \eqref{eqn:
  regularity-condition}.

\section{Infinite towers of conserved charges}
\label{sec: Conserved-charge}

\indent \indent In this section we define three families of
conserved charges on the phase space $\mathscr{F}$, generalizing
existing constructions \cite{Campiglia:2018dyi}. We start by writing
down an ansatz for the field strength and current that is compatible
with our phase space definition \eqref{eqn: Asymp-fields}, \eqref{eqn: gauge-choice-Iplus} and \eqref{eqn:
  regularity-condition}. Consider a power series expansion of the Maxwell
tensor and the current in $1/r^n$ and $\log r/r^n$ near $\NI^+$ of the
form: 
\begin{subequations}
    \label{eqn: F-j-scri+}
    \begin{align}
    \label{eqn: F-scri+-ru}
    F_{ru}^{\rm sd}(r,u,\hat{x})=&\frac{1}{r^2}\sum_{n=0}\frac{1}{r^n}\overset{n}{F}_{ru}(u,\hat{x})+\frac{1}{r^2}\sum_{n=1}\frac{\log r}{r^n}\overset{n,\ln}{F}_{ru}(u,\hat{x})~,\\
    \label{eqn: j-scri+-A}
    j_A(r,u,\hat{x})=&\frac{1}{r^2}\sum_{n=0}\frac{1}{r^n}\overset{n}{j}_{A}(u,\hat{x})+\frac{1}{r^2}\sum_{n=1}\frac{\log r}{r^n}\overset{n,\ln}{j}_{A}(u,\hat{x})~,\\
    \label{eqn: j-scri+-u}
    j_u(r,u,\hat{x})=&\frac{1}{r^2}\sum_{n=0}\frac{1}{r^n}\overset{n}{j}_{u}(u,\hat{x})+\frac{1}{r^2}\sum_{n=1}\frac{\log r}{r^n}\overset{n,\ln}{j}_{u}(u,\hat{x})~,\\
    \label{eqn: j-scri+-r}
    j_r(r,u,\hat{x})=&\frac{1}{r^4}\sum_{n=0}\frac{1}{r^n}\overset{n}{j}_{r}(u,\hat{x})+\frac{1}{r^4}\sum_{n=1}\frac{\log r}{r^n}\overset{n,\ln}{j}_{r}(u,\hat{x})~.
\end{align}
\end{subequations}
Here we suppress the ``$sd$" in the expansion coefficients throughout
for simplicity. The logarithmic terms appear only at subleading
orders $n\geq1$, consistent with the universality of the leading soft
theorem \cite{Weinberg:1965nx}. These terms are necessary because of
the presence of the $1/u$ tails in the initial data \eqref{eqn:
  regularity-condition}. The expansion \eqref{eqn: F-j-scri+}
generalizes the corresponding expansion of \cite{Campiglia:2018dyi}
which contains no logarithmic terms.

Inserting the ansatz \eqref{eqn:
  F-j-scri+} into Eq.\ \eqref{eqn: j-Fru}, we obtain the following
hierarchy of equations:  
\begin{subequations}
\label{eqn: recursive-F-2}
    \begin{align}
    \label{eqn: recursive-F-n-log}
    2n\partial_u \overset{n,\ln}{F_{ru}}(u,\hat{x}) =& -\big(\Delta +n(n-1)\big) \overset{n-1,\ln}{F_{ru}}(u,\hat{x}) -2D^z\overset{n-1,\ln}{j_z}(u,\hat{x})\nonumber\\
    &+n\overset{n-2,\ln}{j_r}(u,\hat{x}) + 2\partial_u \overset{n-1,\ln}{j_r}(u,\hat{x})~,\\
    2n \partial_u \overset{n}{F_{ru}} (u,\hat{x})=& -\big(\Delta +n(n-1)\big)\overset{n-1}{F}_{ru}(u,\hat{x}) -2D^z\overset{n-1}{j_z}(u,\hat{x})\nonumber\\
    &+n\overset{n-2}{j_r}(u,\hat{x})  + 2\partial_u
    \overset{n-1}{j_r}(u,\hat{x})
    +\left(n-\frac{\Delta }{n}\right)\overset{n-1,\ln}{F_{ru}}(u,\hat{x})\nonumber\\
    &-\frac{2}{n} D^z\overset{n-1,\ln}{j_z}(u,\hat{x})+\frac{2}{n}\partial_u\overset{n-1,\ln}{j_r}(u,\hat{x})~,
    \label{eqn: recursive-F-n-pre}
    \end{align}
\end{subequations}
for $n=1,2,\dots$. Here all terms with negative superscripts should be
understood to be zero:  $\overset{-1}{j_r}=\overset{-1,\ln}{j_r}=0$,
as well as $\overset{0,\ln}{j_r}$ and $\overset{0,\ln}{j_z}$.  
The recursive relations \eqref{eqn: recursive-F-2} allow us to solve
for all the coefficients $\overset{n}{F}_{ru}$ and
$\overset{n,\ln}{F_{ru}}$ for all $n\geq1$, given the first coefficients $\overset{0}{F}_{ru}$ and $\overset{0,\ln}{F_{ru}}$ and
boundary conditions for the differential equations \eqref{eqn:
  recursive-F-2} at $u=+\infty$. We discuss explicitly here the first coefficients (which correspond to $n=0$) and the
case $n=1$.

For $n=0$, the leading order fields are given by \cite{Campiglia:2018dyi}
\begin{subequations}
    \label{eqn: hierarchy-ICs}
    \begin{align}
      \label{eqn: hierarchy-ICs1}
        \partial_u \overset{0}{F}_{ru}=&\overset{0}{j}_u-2\partial_u D^{\bar z}\overset{0}{A}_{\bar z}~,\\ 
        \overset{0,\ln}{F_{ru}}=&0~,
    \end{align}
\end{subequations}
where the second equation is from the expansion ansatz \eqref{eqn: F-scri+-ru}. Following Ref.\,\cite{Strominger:2017zoo} we impose the initial condition 
\begin{equation}
    \label{eqn: n0-IC}
    \overset{0}{F}_{ru}(u,\hat{x})\to 0~,\qquad u\to+\infty~,
\end{equation}
which determines a unique solution.

For $n=1$, the relations (\ref{eqn: recursive-F-2}) reduce to
\begin{subequations}
\label{eqn: recursive-F-2a}
    \begin{align}
    \label{eqn: recursive-F-n-loga}
    \partial_u \overset{1,\ln}{F_{ru}}(u,\hat{x}) =& 0,\\
    2 \partial_u \overset{1}{F_{ru}} (u,\hat{x})=&  -  \Delta \overset{0}{F}_{ru}(u,\hat{x}) -2D^z\overset{0}{j_z}(u,\hat{x})
      + 2\partial_u \overset{0}{j_r}(u,\hat{x})~,
    \label{eqn: recursive-F-n-prea}
    \end{align}
\end{subequations}
which imply that $\overset{1,\ln}{F_{ru}}$ is independent of $u$:
\begin{equation}
  \overset{1,\ln}{F_{ru}}(u,\hat{x}) = \overset{1,\ln}{F_{ru}}(\hat{x}).
  \label{eqn: isconstant}
\end{equation}
We next discuss the initial condition at $u=+\infty$ for Eq.\ (\ref{eqn: recursive-F-n-prea}).
In the absence of the $\log$ terms \cite{Campiglia:2018dyi}, it is consistent to impose vanishing initial conditions
\begin{equation}
    \lim_{u\to+\infty}\overset{n}{F}_{ru}(u,\hat{x})=0~
\end{equation}
for all $n \ge 0$.
However, this assumption is no longer appropriate when the $\log$ terms are present, as discussed in \cite{Compere:2025tzr}. Instead, for example, $\overset{1}{F}_{ru}(u)$ diverges like $\log(u)$ as $u\to+\infty$. 

In the free theory, this logarithmic divergence has a fixed form
\cite{Flanagan:2022pmj}. After projecting to a fixed multipole order
\(\ell\), the coefficient of the late-time \(\log u\) term in
    $\overset{1}{F}_{ru}$
is fixed to
be the negative of the coefficient
    $\overset{1,\ln}{F}_{ru}$
of the explicit \(\log r\)
term. Explicitly, there exists a constant $c_\ell$ for all $\ell \ge 1$
such that
\begin{equation}
    \overset{1}{F}_{ru}(u,\hat{x})
    +
    \overset{1,\ln}{F}_{ru}(\hat{x})
    \log\!\left(\frac{u }{2c_\ell}\right)
    \longrightarrow 0~,
    \quad u\to+\infty~.
    \label{eqn: shifted-n1-IC}
\end{equation}
The explicit value of $c_\ell$ can be shown to be
$c_\ell=e^{\alpha_\ell}$, where
\begin{equation}
    \alpha_\ell
    =
    \sum_{q=2}^{\ell+2}
    \frac{
        (-1)^{\ell+q+1}(\ell+q)!\,H_{q-2}
    }{
        (q-1)!(q-2)!(\ell+1)(\ell+2)(\ell+2-q)!
    }~,
    \quad
    H_m=\sum_{k=1}^{m}\frac{1}{k}~,
    \quad H_0=0 ~.
    \label{eqn: alpha-ell-def}
\end{equation}
Thus the total $O(r^{-3})$ contribution in the expansion (\ref{eqn: F-scri+-ru}) can be written as
\begin{equation}
    \frac{1}{r^3}
    \left[
        \overset{1}{F}_{ru}(u,\hat{x})
        +
        \overset{1,\ln}{F}_{ru}(\hat{x})
        \log\!\left(\frac{u }{2c_\ell}\right)
        +
        \overset{1,\ln}{F}_{ru}(\hat{x})
        \log\!\left(\frac{2rc_\ell}{u}\right)
    \right]~.
    \label{eqn: hierarchy-BC-1}
\end{equation}
The first two terms have a vanishing limit at \(u=+\infty\), while the
third logarithmic term depends only on the dimensionless ratio \(r/u\), up to
a finite constant. This third term is invariant under
the dilation \((u,r)\mapsto(\lambda u,\lambda r)\), whereas a term
proportional to $\log(r u)$ would not be.  This invariance in the
limit $u \to \infty$ is a property of the free solutions.

We will assume that the solutions for the interacting theory also obey
the limiting behavior (\ref{eqn: shifted-n1-IC}) as $u \to \infty$, 
the limit to future timelike infinity $i^+$.  The motivation for this
assumption is that one expects the effects of interactions to be unimportant
in arbitrarily small neighborhoods of $i^+$ in the conformal compactification.
This assumption provides a boundary condition for integrating
Eqs.\ (\ref{eqn: recursive-F-2}) for $n=1$ when logarithmic terms are present.

\subsection{Three classes of conserved charges}
\label{subsec: conserved-charges}

\indent\indent By solving the recursive differential equations
\eqref{eqn: recursive-F-2}
and using Eqs.\ (\ref{eqn: EMEoM}) and (\ref{eqn: regularity-condition-B}), 
we find that the field strengths at order $n$ admit the following expansion as $u\to-\infty$: 
\begin{subequations}
    \label{eqn: Fru-n-large-u}
    \begin{align}
        \overset{n}{F}_{ru}(u,\hat{x})=&\sum_{m=0}^{\infty}\frac{u^n}{u^m}\overset{n,m}{F_{ru}}(\hat{x})+\sum_{m=1}^{n}\frac{u^n\log (-u)}{u^{m}}\overset{n,m,\ln}{F_{ru}}(\hat{x})~,
        \label{eqn: asym-F-n}\\
        \overset{n,\ln}{F_{ru}}(u,\hat{x})=&\sum_{m=1}^n \frac{u^n}{u^m}\overset{n,\ln,m}{F_{ru}}(\hat{x})~.
        \label{eqn: asym-F-log-n}
    \end{align}
\end{subequations}
We will define the charges in terms of some of the coefficients in
these expansions. In particular, we are interested in the
$u$-independent terms $\overset{n,\ln,n}{F_{ru}}(\hat{x})$ and
$\overset{n,n}{F_{ru}}(\hat{x})$, and the coefficient of the
$\log(-u)$ term $\overset{n,n,\ln}{F_{ru}}(\hat{x})$. We also define
analogous quantities $\overset{n,\ln,n}{F_{rv}}(\hat{x})$,
$\overset{n,n}{F_{rv}}(\hat{x})$, and
$\overset{n,n,\ln}{F_{rv}}(\hat{x})$ on $\NI^-$.  Those definitions
are similar except for some sign
flips, see appendix \ref{app:
  past-null-infinity} for details.

Authors of \cite{Campiglia:2018dyi} restrict the free data to a
smaller phase space where the derivative $\partial_u\overset{0}{A}_A(u,\hat{x})$ of the free data falls off faster than
$|u|^{-n}$ as $u\to\pm\infty$ for all $n\geq 0$. Then the field
expansion \eqref{eqn: F-scri+-ru} does not need logarithmic terms,
\begin{equation}
    \overset{n,\ln}{F_{ru}}(\hat{x})=0
\end{equation}
for $n \ge 1$, and the logarithmic terms in the expansion \eqref{eqn: Fru-n-large-u} can be omitted
\begin{equation}
    \overset{n,m,\ln}{F_{ru}}(\hat{x})=0~\quad \forall\,m,\,n\geq0.
\end{equation}
The authors of \cite{Campiglia:2018dyi} then define the infinite tower of charges
\begin{equation}
    Q_n^+[\epsilon^+_n]=\int \d^2\theta\sqrt{q}\,\epsilon^+_n(\hat{x})\overset{n,n}{F_{ru}}(\hat{x})~,\quad Q_n^-[\epsilon^-_n]=\int \d^2\theta\sqrt{q}\,\epsilon^-_n(\hat{x})\overset{n,n}{F_{rv}}(\hat{x})
    \label{eqn: old-charges}
\end{equation}
for $n \ge 0$, where $\epsilon^+_n(\hat{x})$ and $\epsilon^-_n(\hat x)$
are large gauge parameters.
These charges are conserved under scattering if the antipodal matching conditions
\begin{equation}
    \overset{n,n}{F_{ru}}(\hat{x})=(-1)^n\overset{n,n}{F_{rv}}(-\hat{x})
    \label{eqn: old-matching}
\end{equation}
hold, together with the identification
\begin{equation}
   \label{eqn: epsilonidentification}
  \epsilon^+_n(\hat{x})=(-1)^n\epsilon^-_n(-\hat{x}).
\end{equation}

However, for our phase space \eqref{eqn: regularity-condition},
$\overset{n,\ln,m}{F_{ru}}(\hat{x})$ and $
\overset{n,m,\ln}{F_{ru}}(\hat{x})$ are non-vanishing and the matching
conditions \eqref{eqn: old-matching} do not hold. We now explain how
to generalize the definitions \eqref{eqn: old-charges} to this more
general context. We decompose the self-dual field strength
$F_{ru}^{\rm sd}(r,u,\hat{x})$ into spherical harmonics:  
\begin{equation}
  \label{eqn: Fsd}
    F_{ru}^{\rm sd}(r,u,\hat{x})\equiv \sum_{L=0}\sum_{M=-L}^{L}F^+_{LM}(r,u)Y_{LM}(\hat{x})~, 
\end{equation}
where the $+$ superscript refers to $\NI^+$. We rewrite the large-$r$ expansion \eqref{eqn: F-scri+-ru} for each spherical harmonic component $F^+_{LM}(r,u)$ as
\begin{equation}
    F^+_{LM}(r,u)=\frac{1}{r^2}\sum_{n=0}\frac{1}{r^n}\overset{n}{F^+}_{LM}(u)+\frac{1}{r^2}\sum_{n=1}\frac{\log r}{r^n}\overset{n,\ln}{F^+}_{LM}(u)~,
\end{equation}
and recast the $u\to-\infty$ expansion \eqref{eqn: Fru-n-large-u} into the form
\begin{equation}
\label{eqn: Fru-LM-expansion}
    \begin{split}
        F^+_{LM}(r,u)=&\sum_{n=0}\left(\frac{u}{r}\right)^{n+2}\sum_{m=0}^{\infty}\left(\frac{1}{u^{m+2}}\overset{n,m~~}{F^+_{LM}}+\frac{\log\left(-u\right)}{u^{m+2}}\overset{n,m,\ln~}{F^+_{LM}}+\frac{\log r}{u^{m+2}}\overset{n,\ln,m~}{F^+_{LM}}\right)~,
    \end{split}
\end{equation}
where $\overset{n,0,\ln~}{F^+_{LM}}$ and
$\overset{n,\ln,0~}{F^+_{LM}}$ should be understood to be zero. We then define
\begin{equation}
    \overset{n,m~~}{\mathscr F^+_{LM}}=\overset{n,m~~}{F^+_{LM}}-a_{nmL}\overset{n,m,\ln~}{F^+_{LM}}-b_{nmL}\overset{n,\ln,m~}{F^+_{LM}}~,
    \label{eqn: F-LM-nm}
\end{equation}
where $a_{nmL}$ and $b_{nmL}$ are some constants which vanish for $m>n$. The values of these constants will be chosen in sec \ref{sec:relating} to simplify the form of the charges. 

We now define a set of asymptotic charges for fixed $\epsilon^+_n(\hat{x})$ as
\begin{subequations}
\label{eqn: conserved charges-future}
    \begin{align}
        \label{eqn: Q+-ln-1}
        \overset{\ln~~}{Q^+_n}[\epsilon_n^+]=&\sum_{L,M}\left( \overset{n,\ln,n}{F^+_{LM}} +\overset{n,n,\ln}{F^+_{LM}}\right)\epsilon_{n,LM}^+~,
        \\
        \label{eqn: Q+-ln-2}
        \overset{\ln~~}{Q^{'+}_n}[\epsilon_n^+]=&\sum_{L,M}\left( \overset{n,\ln,n}{F^+_{LM}} - \overset{n,n,\ln}{F^+_{LM}}\right)\epsilon_{n,LM}^+ ~,
        \\
        \label{eqn: Q+-n}
        Q^+_n[\epsilon^+_n]=&\sum_{L,M}\overset{n,n}{\mathscr F^+_{LM}}\epsilon_{n,LM}^+~,
    \end{align}
\end{subequations}
where 
\begin{equation}
    \epsilon_{n,LM}^+=\int \d^2\theta\sqrt{q}\,\epsilon^+_n(\hat{x})Y_{LM}(\hat{x})~.
\end{equation}
On past null infinity, we similarly define charge aspects
$\overset{n,\ln,n}{F^-_{LM}}$, $\overset{n,n,\ln}{F^-_{LM}}$, and
$\overset{n,n~~}{F^-_{LM}}$ (see appendix \ref{app:
  past-null-infinity}), and define the charges
\begin{subequations}
\label{eqn: conserved charges past}
    \begin{align}
        \label{eqn: Q--ln-1}
        \overset{\ln~~}{Q^-_n}[\epsilon^-]=&\sum_{L,M}\left( \overset{n,\ln,n}{F^-_{LM}} +\overset{n,n,\ln}{F^-_{LM}}\right)\epsilon_{n,LM}^-~,
        \\
        \label{eqn: Q--ln-2}
        \overset{\ln~~}{Q^{'-}_n}[\epsilon^-]=&-\sum_{L,M}\left( \overset{n,\ln,n}{F^-_{LM}} -\overset{n,n,\ln}{F^-_{LM}}\right)\epsilon_{n,LM}^-~,
        \\
        \label{eqn: Q--n}
        Q^-_n[\epsilon^-]=&\sum_{L,M}\overset{n,n}{\mathscr F^-_{LM}}\epsilon_{n,LM}^-~,
    \end{align}
\end{subequations}
where 
\begin{equation}
    \epsilon_{n,LM}^-=\int \d^2\theta\sqrt{q}\,\epsilon^-_n(\hat{x})Y_{LM}(\hat{x})~.
\end{equation}
Note the extra minus sign in the definition \eqref{eqn: Q--ln-2} of $\overset{\ln~~}{Q^{'-}_n}[\epsilon^-]$, compared to \eqref{eqn: Q+-ln-2}.

These charges are conserved in the sense
\begin{equation}
    \overset{\ln~~}{Q^{+}_n}[\epsilon^+]=\overset{\ln~~}{Q^{-}_n}[\epsilon^-]~,\quad \overset{\ln~~}{Q^{'+}_n}[\epsilon^+]=\overset{\ln~~}{Q^{'-}_n}[\epsilon^-]~,\quad Q^{+}_n[\epsilon^+]=Q^{-}_n[\epsilon^-]~,
    \label{eqn: conservation law}
\end{equation}
provided the following matching conditions hold:
\begin{subequations}
    \label{eqn: matching-cond}
    \begin{align}
        \label{eqn: matching-1}
        \overset{n,\ln,n}{F^+_{LM}} +\overset{n,n,\ln}{F^+_{LM}} =& (-1)^{n+L}\left( \overset{n,\ln,n}{F^-_{LM}} +\overset{n,n,\ln}{F^-_{LM}}\right)\\
        \label{eqn: matching-2}
        \overset{n,\ln,n}{F^+_{LM}} -\overset{n,n,\ln}{F^+_{LM}}=&(-1)^{n+L+1}\left( \overset{n,\ln,n}{F^-_{LM}} -\overset{n,n,\ln}{F^-_{LM}}\right)~,\\
        \label{eqn: matching-3}
        \overset{n,n}{\mathscr F^+}_{LM}=&(-1)^{n+L}\overset{n,n}{\mathscr F^-}_{LM}~.
    \end{align}
\end{subequations}
Here we also assume the identification \eqref{eqn: epsilonidentification} and use the parity property of spherical harmonics $Y_{LM}(-\hat{x})=(-1)^LY_{LM}(\hat{x})$. We will derive the matching conditions \eqref{eqn: matching-cond} in
Sec.\ \ref{sec:relating} below, and in doing so will determine the values of the
coefficients $a_{nmL}$ and $b_{nmL}$ that appear in \eqref{eqn: F-LM-nm}.

%

\subsection{Asymptotic charges at first subleading order}
\label{sec:fso}

The case $n=1$ is of special interest because it is the first order at
which the $1/u$ tail modifies the standard subleading soft charge. We
therefore now specialize the analysis of the preceding section to
$n=1$. 

First, from Eq.\ (\ref{eqn: isconstant}) we have that 
$\overset{1,\ln}{F^+_{LM}}$ is independent of $u$ for all $L$ and
$M$.  It now follows from Eqs.\ \eqref{eqn: Fru-LM-expansion}
and \eqref{eqn: F-LM-nm} that  as $u\to-\infty$
\begin{subequations}
    \begin{align}
        \overset{1}{F^+}_{LM}(u)=&u \overset{1,0}{F^+_{LM}}+\log(-u)\overset{1,1,\ln}{F^+_{LM}}+\overset{1,1}{\mathscr F^+_{LM}}\\
            &+a_{11L}\overset{1,1,\ln}{F^+_{LM}}+b_{11L}\overset{1,\ln,1}{F^+_{LM}}+\mO(u^{-1})~,\nonumber\\
        \overset{1,\ln}{F^+}_{LM}(u)=&\overset{1,\ln,1}{F^+_{LM}}~.
    \end{align}
\end{subequations}
In appendix \ref{app: subleading}, we derive the following expressions for the charges \eqref{eqn: conserved charges-future} as functionals of the free data:
\begin{subequations}
    \label{eqn: Q-1}
    \begin{align}
        \label{eqn: Q-ln-1-+}
        \overset{\ln~~}{Q^+_1}[\epsilon^+]=&\int \d^2z\sqrt{q}\,\epsilon^+u^2\partial_u\Delta  D^{\bar z}\overset{0}{A}_{\bar z}\Big|_{\NI^+_+}-\int \d^2z\sqrt{q}\,\epsilon^+u^2\partial_u\Delta D^{\bar z}\overset{0}{A}_{\bar z}\Big|_{\NI^+_-}~,\\
        \label{eqn: Q-ln-2-+}
        \overset{\ln~}{Q^{'+}_1}[\epsilon^+]=&\int \d^2z\sqrt{q}\,\epsilon^+u^2\partial_u\Delta D^{\bar z}\overset{0}{A}_{\bar z}\Big|_{\NI^+_+}+\int \d^2z\sqrt{q}\,\epsilon^+u^2\partial_u\Delta D^{\bar z}\overset{0}{A}_{\bar z}\Big|_{\NI^+_-}~,\\
        \label{eqn: Q-1-+}
        Q_1^+[\epsilon^+]=&\lim_{\Lambda\to\infty}\left[\int_{-\Lambda}^{\Lambda}\d u\int \d^2z\sqrt{q} \,\epsilon^+\left(D^z\overset{0}{j}_z-\frac{u}{2}\Delta \overset{0}{j}_u+u\partial_u\left(\Delta D^{\bar z}\overset{0}{A}_{\bar z}\right)\right)-\overset{\ln~~}{Q^+_1}[\epsilon^+]\log\Lambda\right]\nonumber\\
        &+\sum_{L,M}(a_{11L}-1)\epsilon^+_{1,LM}u^2\partial_u\left(\Delta  D^{\bar z}\overset{0}{A}_{\bar z}\right)_{LM}\Big|_{\NI^+_-}\nonumber\\
        &+\sum_{L,M}\left(1 -b_{11L} + \log(2c_L) \right)\epsilon^+_{1,LM}u^2\partial_u\left(\Delta  D^{\bar z}\overset{0}{A}_{\bar z}\right)_{LM}\Big|_{\NI^+_+}~,
    \end{align}
\end{subequations}
where the $LM$-subscript stands for the $LM$-component, $\NI^+_\pm$ are future/past ends of $\NI^+$, and
$c_L$ is defined in Eq.\ (\ref{eqn: shifted-n1-IC}).
Similar expressions can be obtained for the charges \eqref{eqn: conserved charges past} on $\NI^-$.

Some comments are in order:
\begin{itemize}
    \item The charge \eqref{eqn: Q-ln-1-+} coincides with the
      logarithmic charge defined in Refs.\ \cite{Campiglia:2019wxe,
        Compere:2025tzr, Choi:2024mac, Choi:2024ygx,Hirai:2018ijc} where the source
      field was assumed to be massive. We extend the definition to massless sources. 
    \item The charge \eqref{eqn: Q-ln-2-+} has not been analyzed before, and we conjecture that it is associated with a new asymptotic symmetry. Gravitational counterparts of the corresponding additional tail-sensitive matching structure have recently been developed in Refs. \cite{Compere:2026jmk,Boschetti:2026ogm} and incorporated into a systematic invariant framework of subleading gravitational charges in Ref. \cite{Compere:2026mdj}.
    \item Our charge \eqref{eqn: Q-1-+} agrees with charges previously
      defined \cite{Conde:2016csj,Campiglia:2016hvg,Campiglia:2018dyi}
      when restricted to the smaller phase space used there. In the
      context of our larger phase space \eqref{eqn:
        regularity-condition}, the charges of Refs.\ \cite{Conde:2016csj,Campiglia:2016hvg,Campiglia:2018dyi}
      are divergent, and our definition \eqref{eqn: Q-1-+} provides a
      natural regularisation. Specifically, the limit
      \eqref{eqn: Q-1-+} exists since the last term in the large
      square brackets cancels the possible divergence in the integral. Hence we regard our definition as a natural renormalisation of the subleading soft charge first discovered in \cite{Lysov:2014csa,Kapec:2015ena}. 
    \item We note that if we replace the factor $\log\Lambda$ with $\log\left(\Lambda/\Lambda_0\right)$ in the first line of \eqref{eqn: Q-1-+}, the effect is to make the replacement $Q_1^+\to Q_1^++\log\Lambda_0\overset{\ln~~}{Q^+_1}$. This transformation corresponds to a shift of $a_{11L}$ and $b_{11L}$ by the same constant 
    and can be regarded as a change in coordinates on the manifold of
    charges. Thus, the charge definitions depend on a choice of
    parameterization, but the manifold of charges itself
    \cite{Woodhouse:1992de} does not.
\end{itemize}
We can also write the charges \eqref{eqn: Q-ln-1-+} and \eqref{eqn: Q-ln-2-+} as local integrals over $\NI^+$
\begin{subequations}
    \begin{align}
        \overset{\ln~~}{Q^+_1}[\epsilon^+]=&\int \d u \d^2z\sqrt{q}\,\epsilon^+\partial_u\left(u^2\partial_u\Delta D^{\bar z}\overset{0}{A}_{\bar z}\right)~,\\
        \label{eqn: Q-ln-2-+-local}
        \overset{\ln~}{Q^{'+}_1}[\epsilon^+]=&\int \d u \d^2z\sqrt{q}\,\epsilon^+\partial_u\left(g(u)u^2\partial_u\Delta D^{\bar z}\overset{0}{A}_{\bar z}\right)~,
    \end{align}
\end{subequations}
where $g(u)$ is any smooth function asymptotes to $\pm1$ as $u\to\pm\infty$. Since the integrand is a total derivative, the value of $\overset{\ln~}{Q^{'+}_1}$ is independent of the choice of $g(u)$.

\section{Derivation of matching conditions}
\label{sec: Matching-Condition}
\indent \indent In this section we establish the matching conditions
\eqref{eqn: matching-cond} that underlie the conservation laws
\eqref{eqn: conservation law}. Our approach is an extension of the
strategy used in Ref.\,\cite{Campiglia:2017mua,Campiglia:2018dyi}. First we identify
the charge aspects by taking limits of $F_{ru}^{sd}(r,u,\theta^A)$ and
$F_{rv}^{sd}(r,v,\theta^A)$ to $\NI^+_-$ and $\NI^-_+$. Next we analyze the dynamics of gauge fields near spatial infinity using hyperboloidal coordinates in which the boundary at spatial infinity is conformal to a unit hyperboloid $\mathcal{H}^0$. The Maxwell fields near $i^0$ yield certain data on $\mathcal{H}^0$ defined in \cite{Campiglia:2018dyi}. The
harmonic and antipodal-map framework on $\mathcal{H}^0$ underlying this matching argument has been analyzed systematically for scalar, vector, and tensor fields with arbitrary sources in Ref. \cite{Compere:2025bnf}.

\subsection{Spatial infinity and hyperbolic coordinates}
\label{subsec: Maxwell-eqn-H0}

\indent\indent We now analyze the dynamics of our theory at spatial infinity $i^0$ by taking appropriate limits. Near spatial infinity, it is convenient to use hyperbolic coordinates $(\rho,\tau,\theta^A)\equiv(\rho, y^\alpha)$ defined by
\begin{equation}
    \rho=\sqrt{r^2-t^2}~,\quad \tau=\frac{t}{\sqrt{r^2-t^2}}~.
    \label{eqn: rhotau-ru}
\end{equation}
In these coordinates, Minkowski spacetime is foliated by timelike hyperboloids of constant $\rho$:
\begin{equation}
    \label{eqn: spatial-hyperbolic}
    ds^2=d\rho^2 + \rho^2h_{\alpha\beta}dy^\alpha dy^\beta=d\rho^2 + \rho^2\left[ -\frac{d\tau^2}{1+\tau^2} + (1+\tau^2)q_{AB}d\theta^Ad\theta^B \right]~,
\end{equation}
where $q_{AB}$ is the metric of a unit two-sphere and $h_{\alpha\beta}$ is the metric of the unit hyperboloid $\mathcal{H}^0$. We also introduce the embedding map
\begin{equation}
    Y: \mathcal{H}^0\to M~:\quad y^\alpha=(\tau,\theta^A) \mapsto (\tau, \sqrt{1+\tau^2}\hat{x})~,
\end{equation}
which maps the asymptotic hyperboloid $\mathcal{H}^0$ onto a unit timelike hyperboloid in Minkowski spacetime. 

We are interested in the regime obtained by first taking the limit to $\NI^+$ followed by a limit along $\NI^+$ to $i^0$. Parametrically, this regime is given by
\begin{equation}
    \Big|\frac{u}{r}\Big|\ll1~,\quad \Big|\frac{1}{u}\Big|\ll1~.
\end{equation}
Inverting the coordinate transformation \eqref{eqn: rhotau-ru}, we have 
\begin{equation}
    r=\rho\sqrt{1+\tau^2}~,\quad u=\rho \big( \tau-\sqrt{1+\tau^2} \big)~,
    \label{eqn: ru-rhotau}
\end{equation}
and hence
\begin{equation}
    \frac{1}{u}=-\frac{2\tau}{\rho}\left(1+\mO(\tau^{-2})\right)~,\qquad
    \frac{u}{r}=-\frac{1}{2\tau^2}\left(1+\mO(\tau^{-2})\right)~.
\end{equation}
It follows that the above regime is equivalently described by
\begin{equation}
  \label{eqn: regime}
    \Big|\frac{\tau}{\rho}\Big|\ll 1~,\qquad \Big|\frac{1}{\tau}\Big|\ll 1~.
\end{equation}
This regime can be reached by first taking the blow-up limit \(\rho\to+\infty\) to reach the asymptotic hyperboloid \(\mathcal H^0\), and then sending \(\tau\to+\infty\) to its future boundary.

\subsection{Maxwell's equations near spatial infinity}
\label{subsec: decoupling-Maxwell-Eqn}

\indent \indent In the phase space \eqref{eqn: regularity-condition},
we assume that the scalar field decays at spatial infinity faster than any
inverse power of $\rho$: 
\begin{equation}
    \Phi(\rho,y^\alpha)=o(\rho^{-n})~
    \label{eqn: Phi-fast-decay}
\end{equation}
for all $n \ge 1$.
As a result, the matter current in Eq.\ \eqref{eqn: EMEoM} becomes
negligible in the $\rho\to\infty$ limit and the coupled
Maxwell--scalar system reduces to the source-free Maxwell theory near
spatial infinity. 
Physically, this means that the infrared behavior is entirely captured by the electromagnetic field, while the charged matter does not contribute at the orders relevant for the matching analysis. This is the sense in which one excludes the ``soft limit of charged particles" as proposed in \cite{Campiglia:2018dyi}.

In the analysis below, we will prove the matching conditions
\eqref{eqn: matching-cond} for the real-valued field $F_{ab}$. From
the definition \eqref{eqn: self-dual field} of the self-dual field,
the matching conditions \eqref{eqn: matching-cond} for the complex-valued self-dual field \(F^{\rm sd}_{ab}\) decompose into the real
part for \(F_{ab}\) and imaginary part for the dual field \(\tilde
F_{ab}\). Because in the asymptotic region near spatial infinity the
real field \(F_{ab}\) obeys the source-free Maxwell's equations which are
invariant under the Hodge dual operation, the same matching argument
applies to the dual field. Combining the two real matching conditions
gives the desired matching condition for \(F^{\rm sd}_{ab}\). 

Performing a $3+1$ decomposition of the Maxwell tensor in the hyperbolic
coordinates \eqref{eqn: spatial-hyperbolic}, we end up with a two-form
$F_{\alpha\beta}(\rho,y)$ on $\mathcal{H}^0$ together with a one-form
$F_{\rho\alpha}(\rho,y)$. The source-free Maxwell's equations
then become
\begin{subequations}
    \label{eqn: Maxwell-H0}
    \begin{align}
    \label{eqn: Maxwell-1form}
    \frac{1}{\rho^2}\mD^\alpha F_{\rho\alpha}(\rho,y) =&\,0~,\\
    \label{eqn: Maxwell-2form}
    -\partial_\rho F_{\rho\alpha}(\rho,y) - \frac{1}{\rho}F_{\rho\alpha}(\rho,y) + \frac{1}{\rho^2}\mD^\beta F_{\alpha\beta}(\rho,y) =&\,0~,\\
    \label{eqn: Bianchi-H-2form}
    \mD_{[\alpha}F_{\beta\gamma]}(\rho,y) =&\,0~,\\
    \label{eqn: Bianchi-H-1form}
    \mathcal{D}_\alpha F_{\rho\beta}(\rho,y) - \mathcal{D}_\beta F_{\rho\alpha}(\rho,y) -\partial_\rho F_{\alpha\beta}(\rho,y) =&\,0~,
    \end{align}
\end{subequations}
where $\mD_\alpha$ is the covariant derivative with respect to the hyperbolic metric $h_{\alpha\beta}$. 

We now assume a large-$\rho$ expansion ansatz near spatial infinity,
similar to the expansion \eqref{eqn: F-j-scri+}, of the form
\begin{subequations}
    \label{eqn: F-H0}
    \begin{align}
        \label{eqn: F-H0-ra}
        F_{\rho\alpha}(\rho,y)=&\frac{1}{\rho}\sum_{k=0}^{\infty}\left(\frac{1}{\rho^k}\overset{k}{F}_{\rho\alpha}(y)+\frac{\log\rho}{\rho^k}\overset{k,\ln}{F}_{\rho\alpha}(y)\right)~,\\
        F_{\alpha\beta}(\rho,y)=&\sum_{k=0}^{\infty}\left(\frac{1}{\rho^k}\overset{k}{F}_{\alpha\beta}(y)+\frac{\log\rho}{\rho^k}\overset{k,\ln}{F}_{\alpha\beta}(y)\right)~.
    \end{align}
\end{subequations}
Plugging the ansatz \eqref{eqn: F-H0} into Maxwell's equations \eqref{eqn: Maxwell-H0}, we obtain
\begin{subequations}
    \begin{align}   
        \label{eqn: Div-free-rho-alpha}
        \mD^\alpha\overset{k,\ln}{F}_{\rho\alpha}=\mD^\alpha \overset{k}{F}_{\rho\alpha}=&0~,\\
        k \overset{k,\ln}{F}_{\rho\alpha}+\mD^\beta \overset{k,\ln}{F}_{\alpha\beta}=&0~,\\
        k \overset{k}{F}_{\rho\alpha}-\overset{k,\ln}{F}_{\rho\alpha}+\mD^\beta \overset{k}{F}_{\alpha\beta}=&0~,\\
        \mathcal{D}_{[\alpha}\overset{k,\ln~~~}{F_{\beta\gamma]}}=\mathcal{D}_{[\alpha}\overset{k}{F}_{\beta\gamma]}=&0~,\\
        \mathcal{D}_\alpha \overset{k,\ln}{F_{\rho\beta}}-\mathcal{D}_\beta \overset{k,\ln}{F_{\rho\alpha}}+k\overset{k,\ln}{F_{\alpha\beta}}=&0~,\\
        \mathcal{D}_\alpha \overset{k}{F}_{\rho\beta} -\mathcal{D}_\beta \overset{k}{F}_{\rho\alpha} + k\overset{k}{F}_{\alpha\beta}-\overset{k,\ln}{F_{\alpha\beta}}=&0~.
    \end{align}
\end{subequations}
Eliminating $\overset{k}{F}_{\alpha\beta}$ and
$\overset{k,\ln}{F_{\alpha\beta}}$ then yields
\begin{subequations}
    \label{eqn: eom-rho-tau}
    \begin{align}
        \label{eqn: eom-rho-tau-n-log}
        \left[\mathcal{D}^2+k^2-2 \right]\overset{k,\ln}{F}_{\rho\alpha}(y)=&0~,\\
        \label{eqn: eom-rho-tau-n}
        \left[\mathcal{D}^2+k^2-2 \right]\overset{k}{F}_{\rho\alpha}(y) -2k\overset{k,\ln}{F}_{\rho\alpha}(y)=&0~.
    \end{align}
\end{subequations}
We will restrict attention below to the case $k \ge 1$.
For $k=0$ we have $\overset{0,\ln~~}{\mF_{LM}}(\tau)=0$, and the analysis coincides with that of Ref.\ \cite{Campiglia:2018dyi}.

Now we need to specify the boundary conditions for
$\overset{k}{F}_{\rho\alpha}$ and $\overset{k,\ln}{F_{\rho\alpha}}$
at $\tau\to+\infty$. As argued in \cite{Campiglia:2018dyi}, the
divergence-free condition \eqref{eqn: Div-free-rho-alpha} relates
$\overset{k}{F}_{\rho\tau}$  to
$\overset{k}{F}_{\rho A}$ and  
$\overset{k,\ln}{F_{\rho\tau}}$ to
$\overset{k,\ln}{F_{\rho A}}$: 
\begin{subequations}
  \begin{eqnarray}
    -\left[(1+\tau^2)\partial_\tau + 3\tau\right] \overset{k
      }{F_{\rho\tau}} + \frac{1}{1+\tau^2} D^A\overset{k
      }{F_{\rho A}} &=& 0, \\
      -\left[(1+\tau^2)\partial_\tau + 3\tau\right] \overset{k
      ,\ln}{F_{\rho\tau}} + \frac{1}{1+\tau^2} D^A\overset{k
      ,\ln}{F_{\rho A}} &=& 0.
\end{eqnarray}
  \end{subequations}
So we can focus on the equations for $\overset{k,\ln}{F_{\rho\tau}}$ and $\overset{k}{F}_{\rho\tau}$.

We decompose the coefficients $\overset{k,\ln}{F_{\rho\tau}}$ and $\overset{k}{F}_{\rho\tau}$ into spherical harmonics:
\begin{subequations}
  \label{eqn: 415}
    \begin{align}
        \overset{k,\ln}{F_{\rho\tau}}(\tau,\hat{x})=&\sum_{L=0}\sum_{M=-L}^{L}\overset{k,\ln}{\mF}_{LM}(\tau)Y_{LM}(\hat{x})~,\\
        \overset{k}{F}_{\rho\tau}(\tau,\hat{x})=& \sum_{L=0}\sum_{M=-L}^{L}\overset{k~~}{\mF_{LM}}(\tau)Y_{LM}(\hat{x})~.
    \end{align}
\end{subequations}
In a given $L$-sector, the differential equations \eqref{eqn: eom-rho-tau} take the form
\begin{subequations}
    \label{eqn: FLM}
    \begin{align}
        \label{eqn: FLM-k-ln}
        \left[(1+\tau^2)\partial^2_\tau + 7\tau\partial_\tau + \frac{L(L+1)-6}{1+\tau^2}-(k^2-9)\right]\overset{k,\ln~~}{\mF_{LM}}(\tau)=&~0~,\\
        \label{eqn: FLM-k}
        \left[(1+\tau^2)\partial^2_\tau + 7\tau\partial_\tau + \frac{L(L+1)-6}{1+\tau^2}-(k^2-9)\right]\overset{k}{\mF}_{LM}(\tau)=&~-2k\overset{k,\ln~~}{\mF_{LM}}(\tau)~.
    \end{align}
\end{subequations}

Two independent solutions of the homogeneous equation \eqref{eqn: FLM-k-ln} are given by
\begin{subequations}
    \label{eqn: homo-solution}
    \begin{align}
      \overset{k,\ln~~}{\mF_{LM}}(\tau)=
      \overset{k}{T}_{L,{\rm even}}(\tau)\propto&~\frac{1}{(1+\tau^2)^{\frac{L+3}{2}}}{}_2F_1\left(\frac{k-L}{2},-\frac{k+L}{2};\frac{1}{2};-\tau^2\right)~,\\
      \overset{k,\ln~~}{\mF_{LM}}(\tau)=        \overset{k}{T}_{L,{\rm odd}}(\tau)\propto&~\frac{\tau}{(1+\tau^2)^{\frac{L+3}{2}}}{}_2F_1\left(\frac{k-L+1}{2},\frac{1-k-L}{2};\frac{3}{2};-\tau^2\right)~,
    \end{align}
\end{subequations}
where $_2F_1$ is the Gauss hypergeometric function.
These solutions are eigenfunctions of the time-reversal map $\tau\to-\tau$:
\begin{equation}
    \label{eqn: parity-T}
    \overset{k}{T}_{L,{\rm even}}(-\tau)=\overset{k}{T}_{L,{\rm
        even}}(\tau)~,\quad \overset{k}{T}_{L,{\rm
        odd}}(-\tau)=-\overset{k}{T}_{L,{\rm odd}}(\tau)~.
\end{equation}
We choose the normalizations of 
$\overset{k}{T}_{L,{\rm even}}(\tau)$ and
$\overset{k}{T}_{L,{\rm odd}}(\tau)$  such that the coefficient of the
leading order power of $\tau$ in the limit $\tau \to +\infty$ is one.

The key observation is that the parity of the solution is tied to the
asymptotic behavior of the solution as $\tau \to \pm \infty$. If $L-k$ is odd, then the odd
solution scales as $\tau^{-k-3}$ and the even solution scales as
$\tau^{k-3}$, while 
if $L-k$ is even the
situation is reversed. We denote the solution with $\tau^{k-3}$
scaling as $\overset{k}{T}_{L,{\rm slow}}$, and the solution with
$\tau^{-k-3}$ scaling as $\overset{k}{T}_{L,{\rm fast}}$.
Then we have
\begin{equation}
    \label{eqn: parity-falloff}
    \overset{k}{T}_{L,{\rm slow}}(-\tau)=(-1)^{k-L+1}
    \overset{k}{T}_{L,{\rm slow}}(\tau)~,\quad \overset{k}{T}_{L,{\rm
        fast}}(-\tau)=(-1)^{k-L} \overset{k}{T}_{L,{\rm fast}}(\tau)~.
\end{equation}
As we will show below, the field expansion \eqref{eqn: Fru-n-large-u}
near null infinity implies that $\overset{k,\ln}{\mF_{LM}}(\tau)$
belongs to the fast-decaying branch.  We can therefore write
\begin{equation}
    \label{eqn: FLM-k-ln-tau}
    \overset{k,\ln~}{\mF_{LM}}(\tau)=\overset{k,\ln,0}{\mF_{LM}}\overset{k}{T}_{L,{\rm
        fast}}(\tau)=\overset{k,\ln,0}{\mF_{LM}}\tau^{-k-3}+\dots~,\quad \tau\to+\infty~,
\end{equation}
thereby defining the coefficient $\overset{k,\ln,0}{\mF_{LM}}$.

We now turn to the inhomogeneous equation \eqref{eqn: FLM-k}.
To construct solutions of this equation, 
notice that although the parameter $k$ in the functions \eqref{eqn:
  homo-solution} was originally defined to take only non-negative integer
values, the functions are well-defined
solutions of Eq.\ \eqref{eqn: FLM-k-ln}
for any non-negative real $k$.
Now differentiating Eq.\ (\ref{eqn: FLM-k-ln}) with respect to $k$
and then evaluating at an integer value of $k$
shows that the derivative $-\partial_k \overset{k,\ln}{\mF_{LM}}$ is a
particular solution of Eq.\ (\ref{eqn: FLM-k}).  Using Eq.\ (\ref{eqn:
  FLM-k-ln-tau})
and adding a general homogeneous solution gives 
\begin{equation}
    \label{eqn: T-inhomo}
    \overset{k}{\mF}_{LM}(\tau)=-\overset{k,\ln,0}{\mF_{LM}}\partial_k
    \overset{k}{T}_{L,{\rm fast}}(\tau) +
    \frac{\overset{k,k,\ln}{\mF_{LM}}}{\hat{s}_{k,L}}\overset{k}{T}_{L,{\rm
        slow}}(\tau)+\overset{k,k}{\mF_{LM}}\overset{k}{T}_{L,{\rm fast}}(\tau)~, 
\end{equation}
where 
\begin{equation}
    \hat{s}_{k,L}\equiv \frac{(k+L)!}{2^{2k-1}k!(k-1)!(L-k)!}~.
    \label{eqn: hat-s}
\end{equation}
Equation \eqref{eqn: T-inhomo} defines\footnote{One can modify the definition of the coordinates
$$
\left(\overset{k,\ln,0}{\mF_{LM}}, \overset{k,k,\ln}{\mF_{LM}}, 
     \overset{k,k}{\mF_{LM}} \right)$$
on the space of solutions
by replacing $\partial_k\overset{k}{T}_{L,{\rm fast}}$ in Eq.\ (\ref{eqn: T-inhomo})
with
$\partial_k\overset{k}{T}_{L,{\rm fast}}-\lambda_0
\overset{k}{T}_{L,{\rm fast}}$ where $\lambda_0$ is a constant.
This transformation is related to the non-universality or
$\Lambda_0$-dependence of the
subleading charge \eqref{eqn: Q-1-+} with $\log \Lambda$
replaced by $\log(\Lambda/\Lambda_0)$, discussed in Sec.\ \ref{sec:fso}.}
the coefficients
$\overset{k,k,\ln}{\mF_{LM}}$ and
$\overset{k,k~~}{\mF_{LM}}$.
The particular value (\ref{eqn: hat-s}) of the normalization constant
is chosen to simplify expressions that will be encountered later when
matching to $\NI^\pm$.  Note that the expression (\ref{eqn: hat-s}) 
is well-defined only
for $L \ge k$, but this is sufficient since 
$\overset{k}{\mF}_{LM}$ is
zero for $L<k$ as pointed out in Ref.\ \cite{Campiglia:2018dyi}.

Taking the $\tau\to+\infty$ limit of the solution
\eqref{eqn: T-inhomo} using Eqs.\ (\ref{eqn: homo-solution}) and (\ref{eqn: parity-falloff}), we obtain
\begin{equation}
    \begin{split}
        \label{eqn: FLM-k-tau}
        \overset{k}{\mF}_{LM}(\tau) = &\frac{\overset{k,k,\ln}{\mF_{LM}}-\overset{k,\ln,0}{\mF_{LM}}}{\hat{s}_{k,L}}\sum_{l=0}^{k-1}s_l(k,L)\tau^{k-3-2l}+\overset{k,k,\ln}{\mF_{LM}} \tau^{-k-3}\log\tau\\
        &+ \left(\frac{C_{k,L}}{\hat{s}_{k,L}}\overset{k,\ln,0}{\mF_{LM}}+\frac{D_{k,L}}{\hat{s}_{k,L}}\overset{k,k,\ln}{\mF_{LM}}+ \overset{k,k~~}{\mF_{LM}}\right)\tau^{-k-3}+\mO(\tau^{-k-5}\log\tau)~.
    \end{split}
\end{equation}
Here the coefficients are given by 
\begin{subequations}
    \begin{align}
        s_{l}(k,L)
        &=
        \sum_{r=0}^{l}
        \binom{-\frac{L+3}{2}}{l-r}\,
        (-1)^r\,
        \frac{\left(-\frac{k+L}{2}\right)_r\left(\frac{1-k-L}{2}\right)_r}{(1-k)_r\,r!} \,,
        \qquad 0\le l \le k-1 ~,\\
        C_{k,L}
        &=
        -\sum_{r=0}^{k-1}
        \binom{-\frac{L+3}{2}}{k-r}\,
        (-1)^r\,
        \frac{\left(-\frac{k+L}{2}\right)_r\left(\frac{1-k-L}{2}\right)_r}{(1-k)_r\,r!}~,\\
        D_{k,L}
        &=
        \sum_{r=0}^{k-1}
        \binom{-\frac{L+3}{2}}{k-r}\,
        (-1)^r\,
        \frac{\left(-\frac{k+L}{2}\right)_r\left(\frac{1-k-L}{2}\right)_r}{(1-k)_r\,r!}+\frac{\hat{s}_{k,L}}{2}\Bigl(H_k+2\log 2-2H_{L-k}\Bigr)~,
    \end{align}
\end{subequations}
where $\binom{m}{n}$ is generalized binomial coefficient, $(p)_r$ is the Pochhammer symbol, and $H_n$ is the $n^{th}$ Harmonic number. 



\subsection{Relating the field expansions at null and spatial infinity}
\label{sec:relating}

We now recast the expansion \eqref{eqn: Fru-LM-expansion} in hyperbolic coordinates, with the aim of extracting the large-$\rho$ coefficients of $F_{\rho\tau}(\rho,y)$ for each $L$-sector. From Eq.\,\eqref{eqn: rhotau-ru} we have the relation between $F_{\rho\tau}$ and $F_{ru}$:  
\begin{equation}
    \label{eqn: Frt-Fru}
    F_{\rho\tau}=\frac{\rho}{\sqrt{1+\tau^2}}F_{ru}~.
\end{equation}
We next insert into the left hand side of Eq.\ (\ref{eqn: Frt-Fru}) the expressions
(\ref{eqn: F-H0-ra}) and (\ref{eqn: 415}).  On the right hand side we insert the definitions
(\ref{eqn: Fsd}) -- (\ref{eqn: F-LM-nm}), and eliminate $r$ and $u$ in
favor of $\rho$ and $\tau$ using Eq.\ (\ref{eqn: ru-rhotau}).   
Finally expanding in powers of $1/\rho$ and equating the coefficients
of $1/\rho^{k+1}$ and $\log(\rho)/\rho^{k+1}$ on both sides
yields\footnote{In this section, the various coefficients $F^+_{LM}$
are those of the real field $F_{ru}$, rather than those of the
self-dual field $F^{sd}_{ru}$ as in Sec.\ \ref{sec: Conserved-charge}.  See the discussion
before Eqs.\ (\ref{eqn: Maxwell-H0}) above.}
\begin{subequations}
  \label{eqn: Frt-both}
    \begin{align}
        \label{eqn: Frt-k}
        \overset{k}{\mF}_{LM}(\tau)=&\frac{1}{\sqrt{1+\tau^2}}\frac{1}{(\tau-\sqrt{1+\tau^2})^{k+2}}\nonumber\\
        &\times\left[ \sum_{n=0}^{\infty}\left(\frac{\tau}{\sqrt{1+\tau^2}}-1\right)^{n+2}\left(\overset{n,k}{\mathscr F_{LM}^+}+a_{nkL}\overset{n,k,\ln}{F_{LM}^+}+b_{nkL}\overset{n,\ln,k}{F_{LM}^+}\right)\right.\\
        &\quad \quad +\sum_{n=k}^{\infty} \left(\frac{\tau}{\sqrt{1+\tau^2}}-1\right)^{n+2}\log\left(1+\tau^2-\tau\sqrt{1+\tau^2}\right)\frac{\overset{n,k,\ln}{F_{LM}^+}+\overset{n,\ln,k}{F_{LM}^+}}{2}\nonumber\\
        &\quad \quad\left.+\sum_{n=k}^{\infty} \left(\frac{\tau}{\sqrt{1+\tau^2}}-1\right)^{n+2}\log\left(1-\frac{\tau}{\sqrt{1+\tau^2}}\right)\frac{\overset{n,k,\ln}{F_{LM}^+}-\overset{n,\ln,k}{F_{LM}^+}}{2}\right]~,\nonumber\\
        \label{eqn: Frt-k-log}
        \overset{k,\ln}{\mF_{LM}}(\tau)=&\frac{1}{\sqrt{1+\tau^2}}\frac{1}{(\tau-\sqrt{1+\tau^2})^{k+2}}\sum_{n=k}^{\infty} \left(\frac{\tau}{\sqrt{1+\tau^2}}-1\right)^{n+2}\left( \overset{n,k,\ln}{F_{LM}^+}+\overset{n,\ln,k}{F_{LM}^+}\right)~.
    \end{align}
\end{subequations}
These expressions are valid in the limit $\tau \to \infty$; see the
discussion around Eq.\ (\ref{eqn: regime}).

We now compare these results to the expansions \eqref{eqn: FLM-k-ln-tau} and \eqref{eqn: FLM-k-tau}.
First, the leading $\tau \to \infty$ behavior of $\overset{k,\ln}{\mF}_{LM}$ from (\ref{eqn: Frt-k-log}) is in agreement with
Eq.\ \eqref{eqn: FLM-k-ln-tau}, due to the range of the summation index
$n$.
Comparing coefficients of powers of $\tau$ now yields
\begin{subequations}
    \label{eqn: Frt-Fru-relation-k}
    \begin{align}
        \overset{k,\ln,0}{\mF_{LM}}=& \overset{k,\ln,k}{F^+_{LM}}+\overset{k,k,\ln}{F^+_{LM}}~,\\
        \overset{k,k,\ln}{\mF_{LM}}=& \overset{k,\ln,k}{F^+_{LM}}-\overset{k,k,\ln}{F^+_{LM}}~,\\
        \label{eqn: FLMkk}
        \overset{k,k~~}{\mF_{LM}}=&
        \overset{k,k}{\mathscr F^+_{LM}}~.
    \end{align}
\end{subequations}
Here we choose the values of the coefficients $a_{nkL}$ and $b_{nkL}$
that appear in the definition (\ref{eqn: F-LM-nm}) and the expression (\ref{eqn: Frt-k})
to ensure the validity of Eq.\ (\ref{eqn: FLMkk}).
Similarly, as $\tau\to-\infty$, we have\footnote{See appendix
\ref{app: past-null-infinity} for more details.} 
\begin{subequations}
    \label{eqn: Frt-Fru-relation-k--}
    \begin{align}
        \overset{k,\ln,0}{\mF_{LM}}=& (-1)^{k+L}\left(\overset{k,\ln,k}{F^-_{LM}}+\overset{k,k,\ln}{F^-_{LM}}\right)~,\\
        \overset{k,k,\ln}{\mF_{LM}}=& (-1)^{k+L+1}\left(\overset{k,\ln,k}{F^-_{LM}}-\overset{k,k,\ln}{F^-_{LM}}\right)~,\\
        \label{eqn: FLMkk1}
        \overset{k,k~~}{\mF_{LM}}=&
        (-1)^{k+L}\overset{k,k}{\mathscr F^-_{LM}}~.
    \end{align}
\end{subequations}
Combining the results \eqref{eqn: Frt-Fru-relation-k} and \eqref{eqn:
  Frt-Fru-relation-k--} finally yields the matching conditions \eqref{eqn:
  matching-cond}
that underlie the conservation laws (\ref{eqn: conservation law}).

\subsubsection{Matching conditions at subleading order}

In this subsection we give more details for the case $k=1$, corresponding to the first subleading order. 
In this case we obtain from the $\tau\to+\infty$
limit of Eqs.\ (\ref{eqn: Frt-both}) that
\begin{subequations}
    \begin{align}
        \overset{1,\ln}{\mF_{LM}}(\tau)=& \frac{1}{\tau^4}\left( \overset{1,\ln,1}{F_{LM}^+}+\overset{1,1,\ln}{F_{LM}^+}\right)+\mO\left(\tau^{-6}\right)~,\\
        \overset{1}{\mF}_{LM}(\tau)=& 
        -\frac{2}{\tau^2}\overset{0,1}{F^+_{LM}}+ \frac{\log\tau}{\tau^4}\left( \overset{1,\ln,1}{F_{LM}^+}-\overset{1,1,\ln}{F_{LM}^+}\right)\nonumber\\   
        &+ \frac{1}{\tau^4}\left(\overset{1,1}{\mathscr F^+_{LM}} +\frac{5}{2}\overset{0,1}{F^+_{LM}} +\left(a_{11L}-\log2\right) \overset{1,1,\ln}{F_{LM}^+}+ b_{11L}\overset{1,\ln,1}{F_{LM}^+}\right)\nonumber\\
        &+\mO\left(\tau^{-6}\log\tau\right)~.
    \end{align}
\end{subequations}
We now choose the coefficients $a_{11L}$ and $b_{11L}$ to be 
\begin{subequations}
    \begin{align}
        a_{11L}=&H_{L}-\frac{1}{2}H_1+\frac{1}{L+1}~,\\
        b_{11L}=&\frac{1}{2}H_1-H_{L-1}+\log2~.
    \end{align}
\end{subequations}
Then the matching condition for $k=1$ is given by
\begin{subequations}
    \label{eqn: Frt-Fru-relation-1}
    \begin{align}
         \overset{1,\ln,1}{F^+_{LM}}+\overset{1,1,\ln}{F^+_{LM}}=&(-1)^{L+1}\left(\overset{1,\ln,1}{F^-_{LM}}+\overset{1,1,\ln}{F^-_{LM}}\right)~,\\
        \overset{1,\ln,1}{F^+_{LM}}-\overset{1,1,\ln}{F^+_{LM}}=&(-1)^L\left(\overset{1,\ln,1}{F^-_{LM}}-\overset{1,1,\ln}{F^-_{LM}}\right)~,\\
        \label{eqn: FLM11}
        \overset{1,1}{\mathscr  F^+_{LM}}=&
        (-1)^{L+1}\overset{1,1}{\mathscr F^-_{LM}}~.
    \end{align}
\end{subequations}
These equations generalize the matching conditions derived
in Ref.\ \cite{Campiglia:2018dyi,Campiglia:2019wxe}.


\section{Conclusions and outlook}

In this paper, we enlarged the phase space of \cite{Campiglia:2018dyi} to admit \(1/u\) tails in the free data on \(\NI^\pm\). Within this enlarged phase space, we derived three conserved charges extending the analysis of \cite{Campiglia:2018dyi} in two distinct directions. First, we generalized the logarithmic asymptotic charge to allow massless sources. Second, we uncovered a new matching condition across spatial infinity, leading to a new family of conserved charges. 

Throughout this paper, we assumed, following \cite{Campiglia:2018dyi},
that the scalar field falls off at spatial infinity faster than any
power of the hyperbolic radial coordinate $\rho$. This excludes soft
scalar data at spatial infinity and ensures that the matter source is
asymptotically separated from the radiative Maxwell field, so that the
induced equations on \(\mathcal H^0\) reduce to the source-free
Maxwell's equations. Under this assumption, the inclusion of massive or
fermionic matter should not qualitatively modify the analysis at
$\mathcal{H}^0$. By contrast, extending the construction to
non-abelian gauge theories or gravity would be substantially more
subtle, since the asymptotic equations would then be intrinsically
self-interacting. There has been recent progress in related directions
\cite{Capone:2022gme,Agrawal:2023zea}, including gravitational tail and peeling matching relations \cite{Compere:2026jmk} and an asymptotic derivation of the classical logarithmic soft graviton theorem \cite{Boschetti:2026ogm}. An especially important open
direction is to generalize the definition of the phase space used here
to allow configurations in which both the scalar and Maxwell fields
carry soft data at spatial infinity.  

A natural question is whether the present construction extends to all subleading orders. At tree level, the sub\(^n\)-leading soft charge takes the schematic form \cite{Campiglia:2018dyi}
\begin{equation}
    Q_n^{\rm tree}[\epsilon]\sim \int \d u\d \Omega\, \epsilon(\hat{x})u^n \partial_u \mathcal L\!\left(D^A A_A^{(0)}(u,\hat{x})\right)~,
\end{equation}
where \(\mathcal L\) is a differential operator on \(S^2\). If one allows power-law tails \(D^A A_A^{(0)}\sim u^{-k}\), the naive charge becomes logarithmically divergent at $n=k$ and power-law divergent for $n>k$. Extending the present analysis so as to regularize these divergences at arbitrary $n$ is thus a natural future direction. 

Recent work \cite{Sahoo:2018lxl,Krishna:2023fxg,Karan:2025xks,Banerjee:2026keq} on loop-corrected soft photon theorems suggests a general pattern in which the \(n\)-loop contribution enters at order \(\omega^{n-1}(\log\omega)^n\) in the soft expansion. The corresponding free radiative data are then expected to involve polyhomogeneous tails, schematically $u^{-k-1}\left(\log u\right)^{k}$ at $\NI^+$. In particular, the $\mO\left(\log\omega\right)$ soft theorem is associated with the $1/u$ tail in $\overset{0}{A}_B$ and is one-loop exact, in the sense that higher-loop corrections first arise at parametrically more suppressed orders in the soft expansion. This in turn suggests that it would be worthwhile to enlarge the phase space further so as to admit such polyhomogeneous tails. A complementary enlargement allowing leading logarithmic divergences in
the null data, generated by advanced/retarded radiation with finite energy flux, is analyzed in \cite{Briceno:2025cdu}.

Finally, it would be interesting to understand the algebra generated
by \(\overset{\ln}{Q}_n[\epsilon]\),
\(\overset{\ln}{Q'}_n[\epsilon]\), and \(Q_n[\epsilon]\) on the
enlarged phase space. A better understanding of this phase space,
together with a quantization of the corresponding charge algebra,
should lead to a sharper picture of the infrared structure of gauge
theories.

\section*{Acknowledgements}

This research was supported in part by NSF grant PHY-2409350
and by a Simons Foundation Fellowship.

\appendix
\addtocontents{toc}{\protect\setcounter{tocdepth}{1}}

\section{Maxwell's equations near $\NI^\pm$ to subleading order}
\label{app: subleading}
In this appendix, we derive the explicit expressions \eqref{eqn: Q-1}
for the charges (\ref{eqn: conserved charges-future})
specialized to $n=0$ and $n=1$ 
in terms of the free data $\overset{0}{A_z}$ and $\overset{1}{\phi}$
on $\NI^+$.

We start by solving the relevant hierarchy equations \eqref{eqn: hierarchy-ICs1} and \eqref{eqn: recursive-F-2a}, which we reproduce here for convenience.
\begin{subequations}
    \begin{align}
    \label{eqn: F0-ru-A}
    \partial_u \overset{0}{F}_{ru}=&\overset{0}{j}_u-2\partial_u D^{\bar z}\overset{0}{A}_{\bar z}~,\\
    \label{eqn: hierarchy-1-ln-A}
    2\partial_u \overset{1,\ln}{F_{ru}}=&0~,\\
    2\partial_u \overset{1}{F}_{ru}=&-\Delta  \overset{0}{F}_{ru}-2D^z\overset{0}{j}_z+2\partial_u \overset{0}{j}_r~.
    \label{eqn: hierarchy-1-A}
\end{align}
\end{subequations}

Integrating Eq.\ \eqref{eqn: F0-ru-A} using the initial condition \eqref{eqn:
  n0-IC} yields for the leading-order field strength
\begin{equation}
  \label{eqn: Fruans}
    \overset{0}{F}_{ru}(u,\hat x)=2D^{\bar z}\overset{0,0}{A^+_{\bar z}}(\hat{x})-2D^{\bar z}\overset{0}{A_{\bar z}}(u,\hat{x})-\int^{\infty}_{u}\mathrm{d}u'\,
    \overset{0}{j}_u(u',\hat x)~,
\end{equation}
where we have used Eq.\ (\ref{eqn: regularity-condition-A}).
The current that appears here can be evaluated by 
inserting the field expansion \eqref{eqn: Asymp-fields} into the
definition \eqref{eqn: EMEoM} of the current and comparing with the
expansions (\ref{eqn: F-j-scri+}):
    \begin{align}
        \overset{0}{j}_u=&-ie^2\left(\overset{1}{\phi}\partial_u\overset{1}{\phi^*}-\overset{1}{\phi^*}\partial_u\overset{1}{\phi}\right).
        \label{eqn: j-u-0}
    \end{align}
From our phase space definition \eqref{eqn: regularity-condition}, near the future endpoint, this current satisfies 
\begin{equation}
\label{eq:app-ju-tail-condition-plus}
    \int^{\infty}_{u}\mathrm{d}u'\,
    \overset{0}{j}_u(u',\hat x)
    =
    o(u^{-n}),
    \qquad
    u\to+\infty ,
\end{equation}
for all $n \ge 1$.  Similarly 
near the past endpoint we obtain for all $n \ge 1$
\begin{equation}
\label{eq:app-ju-tail-condition-minus}
    \int^{\infty}_{u}\mathrm{d}u'\,
    \overset{0}{j}_u(u',\hat x)
    =
    \int^{\infty}_{-\infty}\mathrm{d}u'\,
    \overset{0}{j}_u(u',\hat x)
    +
    \mO(u^{-n}),
    \qquad
    u\to-\infty .
\end{equation}
Now combining Eq.\ (\ref{eqn: Fruans}) with the large-$u$ expansion
\eqref{eqn: regularity-condition-A} of $\overset{0}{A}_z$, the
fall-offs (\ref{eq:app-ju-tail-condition-plus}) and
(\ref{eq:app-ju-tail-condition-minus})
and the definition \eqref{eqn: asym-F-n} 
leads to 
\begin{subequations}
    \label{eqn: Fru0-asym-u}
    \begin{align}
        \label{eqn: Fru0-asym-u+}
        \overset{0}{F}_{ru}(u,\hat x)=&-\frac{2D^{\bar z}\overset{0,1}{A_{\bar z}^+}(\hat{x})}{u}+\mO(u^{-2})~,\quad u\to +\infty\\
        \label{eqn: Fru0-asym-u-}
        \overset{0}{F}_{ru}(u,\hat x)=&\overset{0,0}{F}_{ru}(\hat{x})-\frac{2D^{\bar z}\overset{0,1}{A_{\bar z}^-}(\hat{x})}{u}+\mO(u^{-2})~,\quad u\to -\infty~,
    \end{align}
\end{subequations}
where 
\begin{equation}
    \overset{0,0}{F}_{ru}(\hat x)=2D^{\bar z}\overset{0,0}{A^+_{\bar z}}(\hat{x})-2D^{\bar z}\overset{0,0}{A^-_{\bar z}}(\hat{x})-\int^{\infty}_{-\infty}\mathrm{d}u'\,
    \overset{0}{j}_u(u',\hat x)~.
\end{equation}

We next turn to the subleading field strength $\overset{1}{F}_{ru}$.
Inserting Eq.\ \eqref{eqn: Fru0-asym-u+} into Eq.\ (\ref{eqn: hierarchy-1-A}) and integrating both sides yields
\begin{equation}
\label{eqn: F1plus}
  \overset{1}{F}_{ru}(u,\hat x)= \Delta  D^{\bar z}\overset{0,1}{A_{\bar z}^+}(\hat{x})\log u+\mO(1)~,\quad u\to+\infty~.
\end{equation}
In the free theory, we calculated the $\mO(1)$ constant to be $-\Delta 
D^{\bar z}\overset{0,1}{A^+_{\bar z}}\log(2e^{\alpha_\ell})$, where $\alpha_\ell$ is
given in Eq.\ \eqref{eqn: alpha-ell-def}. As discussed after
Eq.\ (\ref{eqn: hierarchy-BC-1}) we assume that this result continues to
be valid in the interacting theory.
Now using the asymptotic form \eqref{eqn: Fru0-asym-u-} of $\overset{0}{F}_{ru}$ near $u\to-\infty$
together with the definition \eqref{eqn: asym-F-n} gives
\begin{equation}
    \overset{1}{F}_{ru}(u,\hat x)=-\frac{u}{2}\Delta \overset{0,0}{F}_{ru}(\hat x) +\Delta  D^{\bar z}\overset{0,1}{A^-_{\bar z}}(\hat x)\log(-u)+\overset{1,1}{F}_{ru}(\hat x)+o(1)~
\end{equation}
for $u \to -\infty$.
Hence we have 
\begin{equation}
    \overset{1,1,\ln}{F_{ru}}(\hat x)=\Delta  D^{\bar z}\overset{0,1}{A^-_{\bar z}}(\hat x)=-\lim_{u\to-\infty}u^2\partial_u \Delta  D^{\bar z}\overset{0}{A}_{\bar z}(u,\hat x)~,
    \label{eqn: A01-}
\end{equation}
and 
\begin{eqnarray}
        \overset{1,1}{F}_{ru}(\hat x)&=&\lim_{u\to-\infty}
        \lim_{\Lambda\to+\infty}\bigg[ \Delta 
          D^{\bar z}\overset{0,1}{A_{\bar z}^+}(\hat{x})\log\left(\frac{\Lambda}{2c_\ell}\right)+\frac{u}{2}\Delta \overset{0,0}{F}_{ru}(\hat
          x) -\Delta  D^{\bar z}\overset{0,1}{A^-_{\bar z}}(\hat x)\log(-u)
           \nonumber \\
        &&-\int^{\Lambda}_u\d u'\left(-\frac{\Delta }{2} \overset{0}{F}_{ru}-D^z\overset{0}{j}_z+\partial_u \overset{0}{j}_r\right)\bigg]~.
\end{eqnarray}
This is in agreement with the expression derived in
\cite{Campiglia:2018dyi}.
Our charge expression \eqref{eqn: Q-1-+} is now obtained by
inserting this result in Eqs.\ (\ref{eqn: F-LM-nm}) and (\ref{eqn: Q+-n}).

Finally we turn to the other two charges.
From the definition \eqref{eqn: asym-F-log-n} we have 
\begin{equation}
    \overset{1,\ln,1}{F_{ru}}(\hat x)=\lim_{u\to-\infty}\overset{1,\ln}{F_{ru}}(u,\hat x).
\end{equation}
However $\overset{1,\ln}{F}_{ru}(u,\hat x)$ is independent of $u$
from Eq.\ (\ref{eqn: hierarchy-1-ln-A}).
Combining this with the $u \to \infty$ boundary condition
\eqref{eqn: shifted-n1-IC} and Eq.\ (\ref{eqn: F1plus}) yields
\begin{equation}
    \overset{1,\ln,1}{F_{ru}}(\hat x)=\lim_{u\to+\infty}\overset{1,\ln}{F}_{ru}(u,\hat x)=-\Delta  D^{\bar z}\overset{0,1}{A^+_{\bar z}}(\hat x)=\lim_{u\to+\infty}u^2\partial_u \Delta  D^{\bar z}\overset{0}{A}_{\bar z}(u,\hat x)~.
    \label{eqn: A01+}
\end{equation}
Combining \eqref{eqn: A01-} and \eqref{eqn: A01+} with the definitions (\ref{eqn: conserved charges-future}) 
yields the charge
expressions \eqref{eqn: Q-ln-1-+}  and \eqref{eqn: Q-ln-2-+}.

\section{Free data and field expansion near past null infinity}
\label{app: past-null-infinity}
In this appendix we provide explicit expressions for the field
expansions near past null infinity. We assume an expansion of the vector potential components in advanced coordinates $(v,r,\theta^A)$ where $v = t + r$, similar to the expansion \eqref{eqn: Asymp-fields}. 
We also expand the self-dual Maxwell tensor component and the currents
\begin{subequations}
\begin{align}
    \label{eqn: F-scri--rv}
    F_{rv}^{\rm sd}(r,v,\hat{x})=&\frac{1}{r^2}\sum_{n=0}\frac{1}{r^n}\overset{n}{F^-_{rv}}(v,\hat{x})+\frac{1}{r^2}\sum_{n=1}\frac{\log r}{r^n}\overset{n,\ln}{F^-_{rv}}(v,\hat{x})~,\\
    \label{eqn: j-scri--A}
    j_A(r,v,\hat{x})=&\frac{1}{r^2}\sum_{n=0}\frac{1}{r^n}\overset{n}{j_{A}^-}(v,\hat{x})+\frac{1}{r^2}\sum_{n=1}\frac{\log r}{r^n}\overset{n,\ln}{j_{A}^-}(v,\hat{x})~,\\
    \label{eqn: j-scri--r}
    j_r(r,v,\hat{x})=&\frac{1}{r^4}\sum_{n=0}\frac{1}{r^n}\overset{n}{j_{r}^-}(v,\hat{x})+\frac{1}{r^4}\sum_{n=1}\frac{\log r}{r^n}\overset{n,\ln}{j_{r}^-}(v,\hat{x})~,
\end{align}
\end{subequations}
where the $-$ superscripts indicate coefficients on $\NI^-$.

As before we assume that the field strengths approach those of a free field as $v\to -\infty$, that is, an analog of
the boundary condition \eqref{eqn: shifted-n1-IC}.  In the limit $v \to \infty$ at
spatial infinity we assume the same kind of expansion as before:
\begin{subequations}
\label{eqn: Frv-n-large-v} 
\begin{align}
        \overset{n}{F}_{rv}(v,\hat{x})=&\sum_{m=0}^{\infty}\frac{v^n}{v^m}\overset{n,m}{F_{rv}}(\hat{x})+\sum_{m=1}^{n}\frac{v^n\log v}{v^{m}}\overset{n,m,\ln}{F_{rv}}(\hat{x})~,
        \label{eqn: asym-F-n--}\\
        \overset{n,\ln}{F_{rv}}(v,\hat{x})=&\sum_{m=1}^n \frac{v^n}{v^m}\overset{n,\ln,m}{F_{rv}}(\hat{x})~.
        \label{eqn: asym-F-log-n--}
\end{align}
\end{subequations}
We also decompose $F_{rv}^{\rm sd}(r,v,\hat{x})$ into spherical harmonics
\begin{equation}
    F_{rv}^{\rm sd}(r,v,\hat{x})\equiv \sum_{L=0}\sum_{M=-L}^{L}F^-_{LM}(r,v)Y_{LM}(\hat{x})~, 
\end{equation}
and rewrite the large-$r$ expansion as
\begin{equation}
    F^-_{LM}(r,v)=\frac{1}{r^2}\sum_{n=0}\frac{1}{r^n}\overset{n}{F^-}_{LM}(v)+\frac{1}{r^2}\sum_{n=1}\frac{\log r}{r^n}\overset{n,\ln}{F^-}_{LM}(v)~.
\end{equation}
We then recast the $v\to+\infty$ expansion \eqref{eqn: Frv-n-large-v} into the form
\begin{equation}
\label{eqn: Frv-LM-expansion}
    \begin{split}
        F^-_{LM}(r,v)=&\sum_{n=0}\left(\frac{v}{r}\right)^{n+2}\sum_{m=0}^{\infty}\left(\frac{1}{v^{m+2}}\overset{n,m~~}{F^-_{LM}}+\frac{\log v}{v^{m+2}}\overset{n,m,\ln~}{F^-_{LM}}+\frac{\log r}{v^{m+2}}\overset{n,\ln,m~}{F^-_{LM}}\right)~,
    \end{split}
\end{equation}
where $\overset{n,0,\ln~}{F^-_{LM}}$ and
$\overset{n,\ln,0~}{F^-_{LM}}$ should be understood to be zero. As
before we define
\begin{equation}
    \overset{n,m~~}{\mathscr F^-_{LM}}=\overset{n,m~~}{F^-_{LM}}-a_{nmL}\overset{n,m,\ln~}{F^-_{LM}}-b_{nmL}\overset{n,\ln,m~}{F^-_{LM}}~.
    \label{eqn: F-LM-nm--}
\end{equation}

By paralleling the derivation of Sec.\ \ref{sec:relating}, we obtain for each $LM$
sector as $\tau\to-\infty$
\begin{subequations}
    \begin{align}
        \label{eqn: Frt-k--}
        \overset{k}{\mF}_{LM}(\tau)=&\frac{1}{\sqrt{1+\tau^2}}\frac{1}{(\tau+\sqrt{1+\tau^2})^{k+2}}\nonumber\\
        &\times\left[ \sum_{m=0}^{\infty}\left(\frac{\tau}{\sqrt{1+\tau^2}}+1\right)^{m+2}\left(\overset{m,k}{\mathscr F_{LM}^-}+a_{mkL}\overset{m,k,\ln}{F_{LM}^-}+b_{mkL}\overset{m,\ln,k}{F_{LM}^-}\right)\right.\\
        &\quad \quad +\sum_{n=k}^{\infty} \left(\frac{\tau}{\sqrt{1+\tau^2}}+1\right)^{n+2}\log\left(1+\tau^2+\tau\sqrt{1+\tau^2}\right)\frac{\overset{n,k,\ln}{F_{LM}^-}+\overset{n,\ln,k}{F_{LM}^-}}{2}\nonumber\\
        &\quad \quad\left.+\sum_{n=k}^{\infty} \left(\frac{\tau}{\sqrt{1+\tau^2}}+1\right)^{n+2}\log\left(1+\frac{\tau}{\sqrt{1+\tau^2}}\right)\frac{\overset{n,k,\ln}{F_{LM}^-}-\overset{n,\ln,k}{F_{LM}^-}}{2}\right]~,\nonumber\\
        \label{eqn: Frt-k-log--}
        \overset{k,\ln}{\mF_{LM}}(\tau)=&\frac{1}{\sqrt{1+\tau^2}}\frac{1}{(\tau+\sqrt{1+\tau^2})^{k+2}}\sum_{n=k}^{\infty} \left(\frac{\tau}{\sqrt{1+\tau^2}}+1\right)^{n+2}\left( \overset{n,k,\ln}{F_{LM}^-}+\overset{n,\ln,k}{F_{LM}^-}\right)~,
    \end{align}
\end{subequations}
together with 
\begin{equation}
    \begin{split}
        \label{eqn: FLM-k-tau--}
        \overset{k}{\mF}_{LM}(\tau) = &\frac{(-1)^{L-k+1}\left(\overset{k,k,\ln}{\mF_{LM}}+\overset{k,\ln,0}{\mF_{LM}}\right)}{\hat{s}_{k,L}}\sum_{l=0}^{k-1}s_l(k,L)(-\tau)^{k-3-2l}\\
        &+(-1)^{L-k+1}\overset{k,k,\ln}{\mF_{LM}} (-\tau)^{-k-3}\log(-\tau)\\
        &+ (-1)^{L-k}\left(\frac{C_{k,L}}{\hat{s}_{k,L}}\overset{k,\ln,0}{\mF_{LM}}-\frac{D_{k,L}}{\hat{s}_{k,L}}\overset{k,k,\ln}{\mF_{LM}}+\overset{k,k~~}{\mF_{LM}}\right)(-\tau)^{-k-3}+\mO(\tau^{-k-5}\log|\tau|)~.
    \end{split}    
\end{equation}
Note that there are some sign flips in Eq.\ \eqref{eqn: FLM-k-tau--}
compared with Eq.\ \eqref{eqn: FLM-k-tau}. Finally comparing
Eqs.\ \eqref{eqn: Frt-k--} and \eqref{eqn: FLM-k-tau--} yields Eq.\ \eqref{eqn:
  Frt-Fru-relation-k--}.



\bibliographystyle{utphys}
\bibliography{references}

\end{document}